\documentclass[pra,twocolumn,showpacs,superscriptaddress,floatfix]{revtex4}
\usepackage{epsfig}
\usepackage{latexsym}
\usepackage{amsmath}
\usepackage{graphics}
\usepackage{bm}

\newcommand\br{\mathbf{r}}

\newcommand\romand{\mathrm{d}}

\begin{document}

\title{On an exact solution of the Thomas-Fermi equation for a trapped
Bose-Einstein condensate with dipole-dipole interactions}
\author{Claudia Eberlein}
\affiliation{Dept of Physics \& Astronomy,
    University of Sussex,
     Falmer, Brighton BN1 9QH, England}
\author{Stefano Giovanazzi}
\affiliation{School of Physics \& Astronomy,
  University of St Andrews, North Haugh, St Andrews KY16 9SS, Scotland}
\author{Duncan H J O'Dell}
\affiliation{Dept of Physics \& Astronomy,
    University of Sussex,
     Falmer, Brighton BN1 9QH, England}

\date{\today}
\begin{abstract}
We derive an exact solution to the Thomas-Fermi equation for a
Bose-Einstein condensate which has dipole-dipole interactions as
well as the usual $s$-wave contact interaction, in a harmonic
trap. Remarkably, despite the non-local anisotropic nature of the
dipolar interaction the solution is an inverted parabola, as in
the pure $s$-wave case, but with a different aspect ratio. Various
properties such as electrostriction and stability are discussed.
\end{abstract}

\pacs{03.75.Hh, 34.20.Cf, 32.80.Qk, 75.80.+q}

 \maketitle

\section{Introduction}
Usually the dominant interatomic interactions in an atomic
Bose-Einstein condensate (BEC) are asymptotically of the van der
Waals (vdW) type, which falls off as $r^{-6}$ and is short-range
in comparison to the de~Broglie wavelength of the atoms. These
interactions can be incorporated into a mean-field description of
the condensate via a delta-function pseudo-potential
\cite{ketterle_varenna,dalfovo99}
\begin{equation} U(\br)=4 \pi \hbar^{2} a_{\mathrm{s}}
\delta(\br) /m \equiv g \delta(\br)
\end{equation}
involving just the $s$-wave scattering length $a_{\mathrm{s}}$ and
atomic mass $m$. The interactions then appear as a cubic
non-linearity in the Gross-Pitaevskii equation \cite{gp-eqn} for
the order parameter $\psi(\br)$ of a trapped zero-temperature BEC
\begin{equation}
\mu \psi(\br)= \left\{-\frac{\hbar^{2}}{2m}\nabla^{2} +
V_{\mathrm{trap}}(\br)+g \vert \psi(\br) \vert^{2} \right\}
\psi(\br) \ , \label{eq:gpe}
\end{equation}
where $\mu$ is the chemical potential. The trapping potential,
$V_{\mathrm{trap}}$, due to a magnetic or optical trap is
typically harmonic, $V_{\mathrm{trap}}=(m/2) [ \omega_{x}^2 x^{2}
+\omega_{y}^2 y^{2}+\omega_{z}^{2}z^{2} ]$.

  The Thomas-Fermi regime for a trapped
BEC is reached when the zero-point kinetic energy is vanishingly
small in comparison to both the potential energy due to the trap
and the interaction energy between atoms \cite{dalfovo99}. Many of
the current atomic BEC experiments \cite{ketterle_varenna} satisfy
these conditions, which tend to hold for condensates containing a
large number of atoms. When the kinetic energy is neglected the
time-independent Gross-Pitaevskii equation (\ref{eq:gpe}) can be
trivially solved for the static condensate density,
\begin{equation}
n(\br)\equiv\vert\psi(\br)
\vert^{2}=[\mu-V_{\mathrm{trap}}(\br)]/g \ \ \mathrm{for}\
n(\br)\geq0,
\end{equation}
and $n(\br)=0$ elsewhere. Thus the density profile is completely
determined by the trapping potential, and in a harmonic trap
$n(\br)$ has an (inverted) parabolic profile and the same aspect
ratio as the trap.

Our aim here is to obtain similar exact results for a BEC in which
dipole-dipole interactions play an important role. Compared to the
vdW interaction the dipolar interaction is long-range and
anisotropic, and consequently these systems can be expected to
exhibit novel behavior including unusual stability properties
\cite{goral00,santos00,lushnikov}, exotic ground states such as
supersolid \cite{giovanazzi2002a,goral2002} and checkerboard
phases \cite{goral2002}, and modified excitation spectra
\cite{yi01,goral2002b}, even to the extent of a roton minimum
\cite{odell03,santos03} in the dispersion relation. We have
recently reported on the exact dynamics of a dipolar BEC in the
Thomas-Fermi limit \cite{odell03b}, including the quadrupole and
monopole shape oscillation frequencies. Here we give the full
derivation of the static solution, which was only stated in
\cite{odell03b}, and investigate stability and electrostriction
(change in volume due to an applied electric field). We also
calculate the dipole-dipole potential outside the boundary of the
condensate, which has a bearing on the distribution of thermal
atoms and on the stability of a dipolar BEC, but also on a lattice
array of dipolar BECs, since the effective giant dipole on each
site is coupled to its neighbors by dipole-dipole interactions
\cite{gross02}.

The long-range part of the interaction between two dipoles separated by
$\br$, and aligned by an external field along a unit vector
$\hat{\mathbf{e}}$, is given by
\begin{equation}
U_{\mathrm{dd}}(\br)= \frac{C_{\mathrm{dd}}}{4 \pi}\,
\hat{{\rm e}}_{i} \hat{{\rm e}}_{j} \frac{\left(\delta_{i
j}- 3 \hat{r}_{i} \hat{r}_{j}\right)}{r^{3}}
\label{eq:staticdipdip}
\end{equation}
where the coupling $C_{\mathrm{dd}}/ (4 \pi)$ depends on the
specific realization. Marinescu and You \cite{marinescu98}
investigated the low-energy scattering of two atoms with
dipole-dipole interactions induced by a static electric field,
$\mathbf{E}=E\hat{\mathbf{e}}$, so that $ C_{\mathrm{dd}}=E^{2}
\alpha^{2}/\epsilon_{0}$. Yi and You \cite{yi00} then went on to
consider a BEC composed of such atoms. Another possible scenario
is permanent magnetic dipoles, $d_{m}$, aligned by an external
magnetic field, $\mathbf{B}=B\hat{\mathbf{e}}$, leading to a
coupling $C_{\mathrm{dd}}=\mu_{0} d_{m}^{2}$. A BEC with magnetic
dipole-dipole interactions was first discussed by G{\'{o}}ral,
Rz{\c{a}}{\.{z}}ewski, and Pfau \cite{goral00}. A measure of the
strength of the long-range dipole-dipole interaction relative to
the s-wave scattering energy is provided by the dimensionless
quantity
\begin{equation}
\varepsilon_{\mathrm{dd}}\equiv \frac{C_{\mathrm{dd}}}{3 g}.
\end{equation}
This definition arises naturally from an analysis of the
frequencies of collective excitations (Bogoliubov spectrum) in a
\emph{homogeneous} dipolar BEC \cite{goral00,giovanazzi2002b}. As
we shall see later on, the BEC is stable as long as $0 \le
\varepsilon_{\mathrm{dd}} < 1$, but loses that stability in the
Thomas-Fermi limit when $\varepsilon_{\mathrm{dd}}
> 1$. Note that in the presence of a strong electric field
the $s$-wave scattering length can be modified
\cite{marinescu98,yi01}, and therefore $g$ and hence
$\varepsilon_{\mathrm{dd}}$ should be treated as effective
quantities when dealing with electrically induced dipoles.

Although the magnetic interaction between two atoms is often
masked by a stronger $s$-wave interaction, two recent proposals
indicate how magnetic dipolar effects can be enhanced, either by
rotating the magnetic field in resonance with a collective
excitation frequency of the system \cite{giovanazzi2002b}, or by
using a Feshbach resonance to reduce the $s$-wave scattering
length \cite{yicondmat}. Other suggestions to realize BECs with
strong dipolar interactions include polar molecules \cite{goral00}
and Rydberg atoms \cite{santos00}. Laser induced (dynamic)
dipole-dipole interactions differ from the static case of
Eq.~(\ref{eq:staticdipdip}) by extra retarded terms, including an
$r^{-1}$ term which is always attractive; they are discussed in
References \cite{odell00} and \cite{melezhik}. In certain
situations this very long-range part of the interaction is
important and can be responsible for unique features such as
self-binding, and plasmon-like collective excitations. Here,
however, we confine ourselves to the static case, which could be
realized by static fields but also by using a laser provided the
atomic separation is considerably smaller than the laser
wavelength.

\section{Thomas-Fermi equation for a
dipolar BEC} \label{sec:TFeqn} Proceeding from the Thomas-Fermi
equation for a static BEC, i.e.\ the time-independent
Gross-Pitaevskii equation without the kinetic energy term, we seek
an exact solution for the density, $n(\br)$, of a condensate with
both dipole-dipole interactions and the usual short-range $s$-wave
scattering in a harmonic trap, which for simplicity (though not
necessity) we take to be cylindrically symmetric
($\omega_{x}=\omega_{y}$). The Thomas-Fermi equation then reads
\begin{equation}
\mu  =  \frac{1}{2}m \left(\omega_{x}^{2} \rho^{2}+\omega_{z}^{2}
z^{2} \right) + g n(\br)  + \Phi_{\mathrm{dd}}(\br)
\label{eq:tfeqn}
\end{equation}
where $\rho^{2}=x^{2}+y^{2}$, and $\Phi_{\mathrm{dd}}(\br)$ is the
mean-field potential due to dipole-dipole interactions
\begin{equation}
\Phi_{\mathrm{dd}}(\br) \equiv \int \romand^{3}r'\
U_{\mathrm{dd}}(\br-\br') n(\br') \; . \label{eq:phidd}
\end{equation}
The intuitive form of Eq.~(\ref{eq:phidd}), which corresponds to
the Born approximation for two-body scattering, has been shown by
Yi and You \cite{yi01} to be accurate for dipole moments of the
order of a Bohr magneton and collisions away from any shape
resonances. Recently Derevianko \cite{derevianko} proposed a more
sophisticated approach to the dipolar scattering problem which
suggests that dipole-dipole interactions can be substantially
larger than previously estimated \cite{yi04b}. However, it appears
that, provided the condensate is not too strongly deformed, the
basic form of Eq.~(\ref{eq:phidd}) with the bare interaction
(\ref{eq:staticdipdip}) remains valid, albeit with a renormalized
coupling $C_{\mathrm{dd}}$ \cite{yi04b}.

The presence of the non-local dipolar mean-field potential
$\Phi_{\mathrm{dd}}(\br)$ means that the Thomas-Fermi Eq.\
(\ref{eq:tfeqn}) is an integral equation and so less trivial to
solve than in the purely local delta-function pseudo-potential
case. However, it is straightforward to demonstrate that this
equation also admits an inverted-parabola as a self-consistent
solution. We begin our analysis with a suggestive re-casting of
the dipole-dipole term using the mathematical identity
\begin{equation}
\frac{\left(\delta_{i j}- 3 \hat{r}_{i}
\hat{r}_{j}\right)}{r^{3}}= - \nabla_{i}
\nabla_{j}\frac{1}{r}-\frac{4 \pi}{3} \delta_{ij} \delta(\br)\;.
\label{eq:relation}
\end{equation}
We can then write
\begin{eqnarray}
\Phi_{\mathrm{dd}}(\br) & = & -C_{\mathrm{dd}}\:
\hat{\mathrm{e}}_{i} \hat{\mathrm{e}}_{j} \left( \nabla_{i}
\nabla_{j} \phi(\br)+\frac{\delta_{ij}}{3}  n(\br)
\right) \label{eq:electrostaticgreen1}\\
\mathrm{with }\ \phi(\br) & \equiv &  \frac{1}{4 \pi} \int
\frac{\romand^{3}r' \;n(\br')}{\vert \br-\br' \vert }\;.
\label{eq:electrostaticgreen2}
\end{eqnarray}
The problem thereby reduces to an analogy with electrostatics, and
we need only calculate the `potential' $\phi(\br)$ arising from
the `static charge' distribution $n(\br)$. In particular,
$\phi(\br)$ given by (\ref{eq:electrostaticgreen2}) must obey
Poisson's equation $\nabla^{2} \phi= - n (\br)$. We adopt the
following inverted-parabola as an ansatz for the density profile
of an $N$-atom condensate
\begin{equation}
n(\br)=n_{0}\left[1-\frac{\rho^{2}}{R_{x}^{2}}-\frac{z^{2}}{R_{z}^{2}}
\right] \ \ \mathrm{for}\ n(\br)\geq 0 \label{eq:densityprofile}
\end{equation}
with radii $R_{x}=R_{y}$ and $R_{z}$, and where the central
density $n_{0}$ is constrained by normalization to be
\begin{equation}
n_{0}=15N/(8 \pi R_{x}^{2}R_{z}) \ .
\end{equation}
Then Poisson's equation is satisfied by an `electrostatic potential' of
the form
\begin{equation}
\phi(\br)= a_{0}+a_{1}\rho^{2}+a_{2}z^{2}+a_{3}\rho^{4}+a_{4}z^{4}
+ a_{5}\rho^{2}z^{2}. \label{eq:generalform}
\end{equation}
However, by Eq.\ (\ref{eq:electrostaticgreen1}), the physical
dipolar contribution $\Phi_{\mathrm{dd}}(\br)$ to the mean-field
potential inside the inverted-parabola BEC
(\ref{eq:densityprofile}) will now itself also be
\emph{parabolic}, just like the potentials due to the harmonic
trap and the local $s$-wave scattering interaction. Thus the
Thomas-Fermi equation (\ref{eq:tfeqn}) contains only parabolic and
constant terms and so, remarkably, just as in the pure $s$-wave
case, in the presence of dipole-dipole interactions a parabolic
density profile is also an exact solution of the Thomas-Fermi
problem in a harmonic trap, although this time we should expect
that the condensate aspect ratio differs from that of the trap. It
remains to determine the coefficients appearing in
(\ref{eq:generalform}) and in (\ref{eq:densityprofile}) and adjust
them in such a way that the Thomas-Fermi equation is satisfied. To
this end we shall evaluate the integral
(\ref{eq:electrostaticgreen2}) for a density of the form
(\ref{eq:densityprofile}). This is an arduous task because the
domain of integration is bounded by and has the symmetry of a
spheroid or, in the general case, even of an ellipsoid.
Calculating the integral is possible only if one takes explicit
account of this symmetry, and we shall demonstrate two independent
ways of doing that. One is to transform into spheroidal
coordinates, use the known Green's function of Poisson's equation
in these coordinates \cite{morse}, and subsequently transform back
into Cartesian coordinates. The other is to start from basics and
integrate over successive thin ellipsoidal shells. While the
former approach is also quite involved it is simpler than the
latter. However, if we were to drop our simplifying assumption of
cylindrical symmetry, the second approach is the only workable as
the general solution of Poisson's equation in general ellipsoidal
coordinates is unmanageably complicated because the separation
constants do not separate in these coordinates
\cite[p.~757]{morse}. The second approach is presented in appendix
\ref{appendixA}.

\section{Green's function in spheroidal coordinates}
We now demonstrate the Green's function approach. For prolate
spheroidal coordinates $(\xi,\eta,\varphi)$ we have
$x=q\sqrt{(\xi^2-1)(1-\eta^2)}\cos\varphi$,
$y=q\sqrt{(\xi^2-1)(1-\eta^2)}\sin\varphi$, $z=q\xi\eta$. Surfaces
of constant $\xi$ are confocal spheroids whose eccentricity is
$1/\xi$, and $\xi$ runs between 1 and $\infty$.  Surfaces of
constant $\eta$ are confocal two-sheet hyperboloids of revolution,
and $\eta$ runs between -1 and 1. For $R_{z}>R_{x}$ the boundary
of the density profile (\ref{eq:densityprofile}) is a prolate
(cigar-like) spheroid with semimajor axis $R_{z}$, semiminor axis
$R_{x}$, and eccentricity $\sqrt{1-R_{x}^{2}/R_{z}^{2}}$. To make
the spheroidal coordinate system confocal to that boundary we need
to choose the scaling constant $q=\sqrt{R_{z}^{2}-R_{x}^{2}}$.
Then we can use the Green's function in prolate spheroidal
coordinates \cite{morse} to write the potential
(\ref{eq:electrostaticgreen2}) as
\begin{eqnarray}
\phi(\xi,\eta,\varphi)  & = &
\frac{R_{z}^{2}-R_{x}^{2}}{2}\left[\int_{1}^{\xi}\romand\xi'
\int_{-1}^{1}\romand\eta'({\xi'}^{2}-{\eta'}^{2})\: n(\xi',\eta')
\right.\nonumber\\
&&\times\sum_{\ell=0}^{\infty}(2\ell +1)
P_\ell(\eta)P_\ell(\eta')Q_\ell(\xi)P_\ell(\xi')\nonumber\\
&&+\int_{\xi}^{1/\sqrt{1-R_{x}^{2}/R_{z}^{2}}}\romand\xi'
\int_{-1}^1\romand\eta'({\xi'}^2-{\eta'}^2)\: n(\xi',\eta')\nonumber\\
&&\left.\times\sum_{\ell=0}^{\infty}(2\ell +1)
P_\ell(\eta)P_\ell(\eta')P_\ell(\xi)Q_\ell(\xi')\right]\;,\nonumber
\end{eqnarray}
where $P_\ell$ are Legendre functions of the first and $Q_\ell$ of
the second kind. Since $n(\br)$ is quadratic in $x$, $y$, and $z$
it is quadratic in $\xi$ and $\eta$, and all integrals in the
above expression are elementary. Performing the $\eta'$
integration first we see that the only contributing $\ell$ are 0,
2, and 4. To re-express the result for $\phi(\xi,\eta,\varphi)$ in
Cartesian coordinates we need to make the substitutions
$\xi=(r_1+r_2)/(2q)$ and $\eta=(r_1-r_2)/(2q)$ with
$r_1=[x^2+y^2+(z+q)^2]^{1/2}$ and $r_2=[x^2+y^2+(z-q)^2]^{1/2}$.
We thereby obtain a `potential' of the form predicted by Eq.\
(\ref{eq:generalform}):
\begin{eqnarray}
\phi(\br)&\!\!\!= &\!\!\!\frac{n_{0}R_{x}^{2}}{192 (1-\kappa^{2})^{2}}\bigg\{
24\Xi (1-\kappa^2)^2 \nonumber\\
&&+48 (1-\kappa^2)(2-\Xi) \left(\frac{z}{R_{z}}\right)^2 \nonumber\\
&&-24 (1-\kappa^2)(2- \kappa^2 \Xi) \left(\frac{\rho}{R_{x}}\right)^{2} \nonumber\\
&&+8 (2\kappa^{2}-8+3  \Xi)\left(\frac{z}{R_{z}}\right)^4 \nonumber\\
&&+3[2(2-5\kappa^{2})+3 \kappa^4 \Xi ] \left(\frac{\rho}{R_{x}}\right)^{4} \nonumber\\
&& +24(2+4\kappa^{2}-3 \kappa^{2} \Xi) \left(\frac{\rho}{R_{x}}\right)^{2}
\left(\frac{z}{R_{z}}\right)^2\bigg\} \label{eq:resultphi}
\end{eqnarray}
where $\kappa \equiv R_{x}/R_{z}$ is the aspect ratio of the BEC
and
\begin{equation}
\Xi \equiv  \frac{1}{\sqrt{1-\kappa^{2}}}\ln\frac{1
+\sqrt{1-\kappa^{2}}}{1-\sqrt{1-\kappa^{2}}} \quad \mathrm{for}
\quad \kappa<1 \quad \mathrm{(prolate)}.
\end{equation}

\begin{figure}[t]
\vspace*{-4mm}
\begin{center}
\centerline{\epsfig{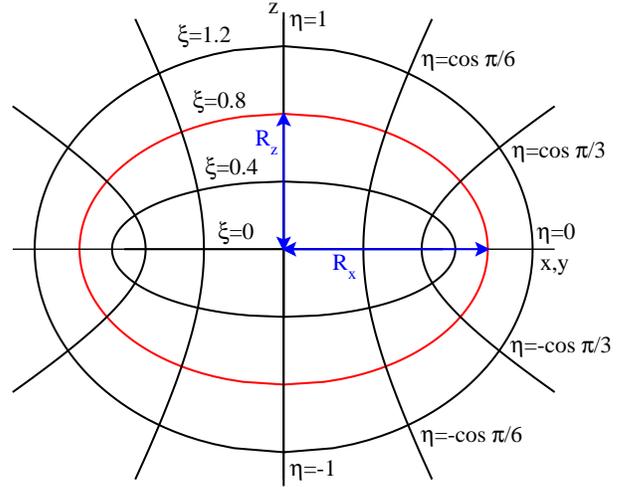}}
\end{center}
\vspace*{-5mm} \caption{An illustration of oblate spheroidal coordinates.
Surfaces of constant $\xi$ are confocal spheroids, with $\xi=0$ being a flat
disk of radius $q$ and $\xi\rightarrow\infty$ an infinitely large
sphere. Surfaces of constant $|\eta|$ are one-sheet hyperboloids
(narrow-waisted reels); $\eta$ changes sign with $z$. The surface described
by $\eta=0$ is the $x y$ plane with a circular hole of radius $q$, and
$|\eta|=1$ is the $z$ axis.}
\label{fig:coordsys}
\end{figure}
If $R_{x}>R_{z}$ then the boundary of the density profile
(\ref{eq:densityprofile}) is an oblate (pancake-like) spheroid, and we have
to use oblate spheroidal coordinates
$x=q\sqrt{(\xi^2+1)(1-\eta^2)}\cos\varphi$,
$y=q\sqrt{(\xi^2+1)(1-\eta^2)}\sin\varphi$, $z=q\xi\eta$. Surfaces of
constant $\xi$ are again confocal spheroids but now with eccentricity
$1/\sqrt{\xi+1}$, and $\xi$ running between 0 and $\infty$. An illustration
of oblate spheroidal coordinates is given in Fig.~\ref{fig:coordsys}. We
have to choose $q=\sqrt{R_{x}^{2}-R_{z}^{2}}$ to make the coordinate system
confocal to the boundary of $n(\br)$. Using the Green's function in oblate
spheroidal coordinates
\cite{morse,mistake} we find for the potential
\begin{eqnarray}
\phi(\xi,\eta,\varphi)& = &\frac{R_{x}^{2}-R_{z}^{2}}{2}
\left[\int_{0}^{\xi}\romand\xi'
\int_{-1}^1\romand\eta'({\xi'}^2+{\eta'}^2)\: n(\xi',\eta') \right.\nonumber\\
&&\times\ {\rm i}\sum_{\ell=0}^{\infty}(2\ell +1)
P_\ell(\eta)P_\ell(\eta')Q_\ell({\rm i}\xi)P_\ell({\rm i}\xi')\nonumber\\
&&+\int_{\xi}^{1/\sqrt{R_{x}^{2}/R_{z}^{2}-1}}\romand\xi'
\int_{-1}^1\romand\eta'({\xi'}^2+{\eta'}^2)\: n(\xi',\eta')\nonumber\\
&&\left.\times\ {\rm i}\sum_{\ell=0}^{\infty}(2\ell +1)
P_\ell(\eta)P_\ell(\eta')P_\ell({\rm i}\xi)Q_\ell({\rm
i}\xi')\right]\;. \nonumber
\end{eqnarray}
To return to Cartesian coordinates we need to make the
substitutions $\xi=(\sqrt{x^2+y^2+(z+{\rm
i}q)^2}+\sqrt{x^2+y^2+(z-{\rm i}q)^2})/(2q)$ and
$\eta=(\sqrt{x^2+y^2+(z+{\rm i}q)^2}-\sqrt{x^2+y^2+(z-{\rm
i}q)^2}) /(2{\rm i}q)$. Then we find that the result for the
potential is the same as in Eq.~(\ref{eq:resultphi}) but with
\begin{equation}
\Xi  \equiv  \frac{2}{\sqrt{\kappa^{2}-1}}\arctan
\sqrt{\kappa^{2}-1} \quad \mathrm{for} \quad \kappa>1 \quad
\mathrm{(oblate)}\;.
\end{equation}
The prolate and oblate cases are of course connected by analytic
continuation, which however cannot be used to determine one from
the other because of the ambiguity of sheets in the complex plane.

In order to simplify the expressions that will follow, and in
order to conform with existing notation in the literature
\cite{yi01,giovanazzi03}, rather than working with the function
$\Xi(\kappa)$, we shall work instead with $f(\kappa)$:
\begin{equation}
f(\kappa) \equiv \frac{2+\kappa^{2}[4-3
\Xi(\kappa)]}{2(1-\kappa^{2})}.
\end{equation}
$f(\kappa)$ is a monotonically decreasing function of $\kappa$,
having values in the range $1 \ge f(\kappa) \ge -2$, passing
through zero at $\kappa=1$.

\section{Solution of the Thomas-Fermi equation}
In this paper we shall take the external field to be along the
$z$-axis. Then the result (\ref{eq:resultphi}) for the
`electrostatic potential' $\phi(\br)$ yields, by virtue of Eq.\
(\ref{eq:electrostaticgreen1}), a parabolic dipolar potential
\begin{eqnarray}
\Phi_{\mathrm{dd}} = \frac{ n_0
C_{\mathrm{dd}}}{3}\left[\frac{\rho^2}{ R_{x}^2}-\frac{2 z^2}{
R_{z}^2}-f \left( \kappa \right)\left(1-\frac{3}{2}\frac{\rho^2 -
2 z^2}{ R_{x}^2- R_{z}^2}\right) \right]\label{phiddinside}
\end{eqnarray}
which is valid inside the condensate (the potential outside the
condensate boundary will be discussed in the next section).
Substituting  $\Phi_{\mathrm{dd}}(\br)$ into the Thomas-Fermi Eq.\
(\ref{eq:tfeqn}) and comparing the coefficients of $\rho^{2}$,
$z^{2}$, and 1, yields three coupled equations. The first
equation, due to the constant terms, gives the chemical potential
\begin{equation}
\mu=g n_{0} \left[1-\varepsilon_{\mathrm{dd}}f(\kappa) \right].
\label{eq:chemical_potential}
\end{equation}
This equation indicates that, all other things being equal, the
effect of dipole-dipole interactions is to lower the chemical
potential (which is proportional to the mean-field energy per
particle) of a prolate ($\kappa < 1$) condensate, whilst raising
that of an oblate ($\kappa > 1$) condensate.
\begin{figure}[t]
\vspace*{-4mm}
\begin{center}
\centerline{\epsfig{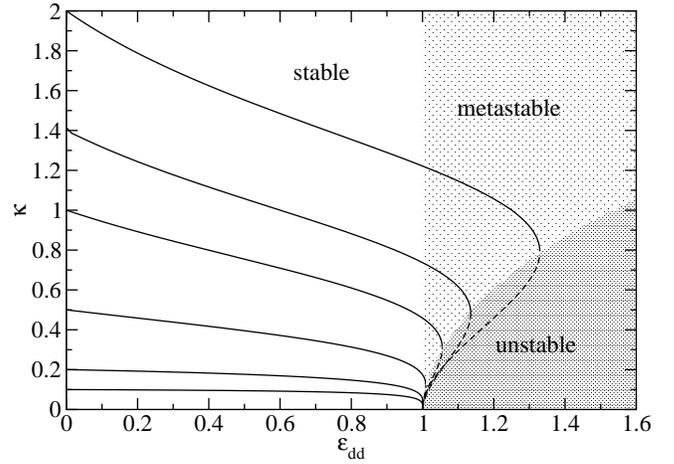}}
\end{center}
\vspace*{-5mm} \caption{Aspect ratio of the condensate as a
function of the dipole-dipole to $s$-wave coupling ratio
$\varepsilon_{\mathrm{dd}}$. Each line is for a different trap aspect ratio
$\gamma$, which can be read off by noting that
$\kappa(\varepsilon_{dd}=0)=\gamma$. When $0 < \kappa < 1$ the condensate is
prolate, for $\kappa > 1$ it is oblate. Likewise, when $0<\gamma <1$ the
trap is prolate, and when $\gamma > 1$ the trap is oblate. Dashed lines
indicate unstable branches.} \label{fig:aspect}
\end{figure}
The radii $R_{x}$($=R_{y}$) and $R_{z}$ of the exact parabolic
solution (\ref{eq:densityprofile}) are obtained from the
coefficients of $\rho^{2}$ and $z^{2}$. We find
\begin{eqnarray}
R_{x}=R_{y}=\left[\frac{15 g N \kappa}{4 \pi m \omega_{x}^{2}}
\left\{1+  \varepsilon_{\mathrm{dd}} \left( \frac{3}{2}
\frac{\kappa^{2} f(\kappa)}{1-\kappa^{2}}-1 \right)
 \right\} \right]^{1/5} \label{eq:Rxsol}
\end{eqnarray}
and $R_{z}=R_{x}/\kappa$. The aspect ratio $\kappa$ is determined
by solving a transcendental equation
\begin{equation}
3 \kappa^{2} \varepsilon_{\mathrm{dd}} \left[
\left(\frac{\gamma^{2}}{2}+1
\right)\frac{f(\kappa)}{1-\kappa^{2}}-1 \right] + \left(
\varepsilon_{\mathrm{dd}}-1 \right)
\left(\kappa^{2}-\gamma^{2}\right)  =0 \label{eq:transcendental}
\end{equation}
where $\gamma=\omega_{z}/\omega_{x}$ is the ratio of the harmonic
trapping frequencies. In fact, a property such as the aspect ratio
is insensitive to the details of the density profile and Eq.\
(\ref{eq:transcendental}) has been obtained previously from a
Gaussian variational ansatz for the density
\cite{yi01,giovanazzi03}. Figures \ref{fig:aspect} and
\ref{fig:aspect2} show examples of the dependence of $\kappa$ upon
$\varepsilon_{\mathrm{dd}}$ for oblate, spherical and prolate
traps. The effect of dipole-dipole forces polarized along the
$z$-axis is to make the condensate more cigar-shaped along $z$.
For an oblate trap ($\gamma >1$) the BEC becomes exactly spherical
when $\varepsilon_{\mathrm{dd}}
=(5/2)(\gamma^{2}-1)/(\gamma^{2}+2)$.

\begin{figure}[t]
\vspace*{-4mm}
\begin{center}
\centerline{\epsfig{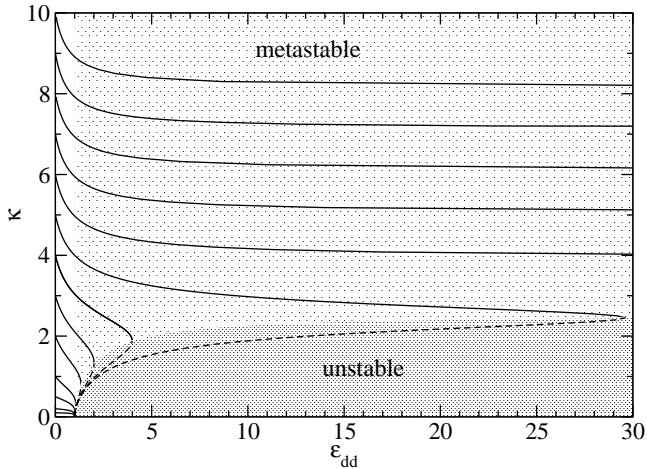}}
\end{center}
\vspace*{-9mm} \caption{Aspect ratio of the condensate as a
function of $\varepsilon_{\mathrm{dd}}$.  This is an expanded
version of Fig.\ \ref{fig:aspect}, illustrating the surprising
result that beyond a certain critical oblateness of the trap
($\gamma
>\gamma_{\mathrm{crit}} = 5.1701$) the system is metastable to
\emph{scaling} deformations, even for arbitrarily high values of
the dipolar interaction strength (but is not necessarily stable to
other types of perturbations such as phonons---see later). At the
boundary value $\gamma=\gamma_{\mathrm{crit}}$, one finds $\kappa
\rightarrow 2.5501$ for $\varepsilon_{\mathrm{dd}}
\rightarrow\infty$.} \label{fig:aspect2}
\end{figure}

 In order to illustrate the static properties
of the Thomas-Fermi solution for a dipolar BEC we imagine an
experiment with a large fixed number of atoms, $N$, in a trap set
to a particular aspect ratio, $\gamma$, where the value of
$\varepsilon_{\mathrm{dd}}$ is adiabatically increased from zero.
For electrically induced dipoles this would involve increasing the
electric field, whereas for magnetic dipoles one could either
rotate the external magnetic field, gradually changing the angle
of rotation \cite{giovanazzi2002b}, or reduce the $s$-wave
scattering length using a Feshbach resonance. The system then
follows one of the curves shown in Figures \ref{fig:aspect} and
\ref{fig:aspect2}. In the absence of the external field, when
$\varepsilon_{\mathrm{dd}}=0$, the condensate aspect ratio matches
that of the trap, $\kappa=\gamma$. When the dipole-dipole
interactions are switched on the condensate becomes more prolate
than the trap and one always has $\kappa < \gamma$. As long as $0
\le \varepsilon_{\mathrm{dd}} <1$, the transcendental equation
(\ref{eq:transcendental}) has a single solution, $\kappa$, for any
choice of trap, $\gamma$. The behavior for
$\varepsilon_{\mathrm{dd}} >1$ is more complicated and requires an
analysis of the stability properties of a dipolar BEC, which we
give in the next but one section.

\section{Dipolar potential outside the BEC} \label{sec:outside}
For a variety of applications it is very useful to know the
potential outside a dipolar condensate. For example, in an array
of dipolar BECs on a lattice the condensate at each lattice site
can behave as a single mesoscopic spin \cite{gross02}. In order to
determine the spin-spin coupling between sites one needs to know
the external potential generated by each condensate. In Section
\ref{subsec:saturn} below we shall see that the knowledge of the
outside potential also gives insight into the problem of the
stability of a single dipolar condensate.

In order to calculate $\Phi_{\mathrm{dd}}({\bf r})$ in Eq.~(\ref{eq:phidd})
outside the condensate, we again use relation (\ref{eq:relation}) and write
the outside dipole-dipole potential in the same way as in
Eq.~(\ref{eq:electrostaticgreen1}), except that the term with $\delta_{ij}$
does not arise because $n({\bf r})$ is obviously zero outside the
condensate. Using the Green's function in prolate spheroidal coordinates we
find for the potential (\ref{eq:electrostaticgreen2}) outside
\begin{eqnarray}
\phi(\xi,\eta,\varphi)& = &
\frac{R_{z}^{2}-R_{x}^{2}}{2}\left[\int_{1}^{1/\sqrt{1-\kappa^{2}}}
\!\!\romand\xi'\!\int_{-1}^{1}\romand\eta'({\xi'}^{2}-{\eta'}^{2})
\right.\nonumber\\&&\hspace*{-6mm}\left.\times
 n(\xi',\eta')
\sum_{\ell=0}^{\infty}(2\ell +1)
P_\ell(\eta)P_\ell(\eta')Q_\ell(\xi)P_\ell(\xi')\right]\;,\nonumber
\end{eqnarray}
where, as before, only $\ell=0,2,4$ actually contribute to the
sum. In the oblate case a similar formula applies, with just $\xi$
and $\xi'$ replaced by i$\xi$ and i$\xi'$ and the $\xi'$
integration running from 0 to $1/\sqrt{\kappa^{2}-1}$. We obtain
\begin{eqnarray}
&&\hspace*{-8mm} \Phi_{\mathrm{dd}}^{\rm (outside)}({\bf r})=
-\frac{3g\varepsilon_{\rm
dd}n_0\kappa^2}{4(1-\kappa^2)^{3/2}}\bigg\{ 6\xi(1-3\eta^2)
\label{eq:outprolate}\\ &&
+\left.\left[9\xi^2\eta^2-3(\xi^2+\eta^2)+1\right] \ln
\frac{\xi+1}{\xi-1} \right\} \nonumber
\end{eqnarray}
in the prolate case, and
\begin{eqnarray}
&&\hspace*{-8mm} \Phi_{\mathrm{dd}}^{\rm (outside)}({\bf r})=
-\frac{3g\varepsilon_{\rm
dd}n_0\kappa^2}{4(\kappa^2-1)^{3/2}}\big\{ 6\xi(1-3\eta^2)
\label{eq:outoblate}\\ &&\hspace*{-8mm}
+\left.\left[9\xi^2\eta^2-3(\xi^2-\eta^2)-1\right]
(\pi-2\arctan\xi) \right\} \nonumber
\end{eqnarray}
in the oblate case. These expressions can easily be converted back
from prolate or oblate spheroidal coordinates $(\xi,\eta)$ into
Cartesian coordinates, as was described above, by substituting
$\xi=(\sqrt{x^2+y^2+(z+q)^2}+\sqrt{x^2+y^2+(z-q)^2})/(2q)$ and
$\eta=(\sqrt{x^2+y^2+(z+q)^2}-\sqrt{x^2+y^2+(z-q)^2}) /(2q)$
for prolate spheroidal coordinates, and
$\xi=(\sqrt{x^2+y^2+(z+{\rm
i}q)^2}+\sqrt{x^2+y^2+(z-{\rm i}q)^2})/(2q)$ and
$\eta=(\sqrt{x^2+y^2+(z+{\rm i}q)^2}-\sqrt{x^2+y^2+(z-{\rm
i}q)^2}) /(2{\rm i}q)$ for oblate spheroidal coordinates.

While easy to obtain, the expression for
$\Phi_{\mathrm{dd}}^{\mathrm{(outside)}}({\bf r})$ is rather lengthy and
unwieldy in Cartesian coordinates. Thus, if one has to work in Cartesian
coordinates, one may prefer the asymptotic expression for the dipole-dipole
interaction potential at large
distances $r=(\rho^2+z^2)^{1/2}$, which is approximately
\begin{eqnarray}
\Phi_{\mathrm{dd}}^{\mathrm{(outside)}}({\bf r}) \simeq
C_{\mathrm{dd}} \frac{N}{4 \pi r^{3}}\left[
\left( 1-\frac{3z^2}{r^2} \right)\right.\hspace*{20mm}\nonumber\\
\left. + \frac{R_{x}^{2}-R_{z}^{2}}{r^2} \left( \frac{9}{14} -
\frac{45}{7} \frac{z^2}{r^2} + \frac{15}{2}\frac{z^4}{r^4}\right)
+ O \left(\frac{R_{x},R_{z}}{r}\right)^4\right] \;,
\label{eq:phiddoutside}
\end{eqnarray}
and holds for both prolate and oblate condensates. It turns out
that this asymptotic expression serves remarkably well even quite
close to the condensate. We give a derivation from integration
over thin ellipsoidal shells in Appendix B. Note that equation
(\ref{eq:phiddoutside}) says that to a first approximation, when
seen from outside the dipolar condensate behaves like a single
giant dipole of N-times the single-atom dipole magnitude. The
higher multipoles depend on the shape of the BEC.

\section{Stability of a dipolar BEC} \label{sec:stability}
The partially attractive nature of the dipole-dipole interaction
(\ref{eq:staticdipdip}) has been widely predicted
\cite{goral00,santos00,yi01,lushnikov} to lead to a collapse of
the BEC when the dipolar interaction strength exceeds a certain
critical value, $\varepsilon_{\mathrm{dd}}^{\mathrm{crit}}$. In
the Thomas-Fermi limit $\varepsilon_{\mathrm{dd}}^{\mathrm{crit}}$
depends only upon the trap aspect ratio $\gamma$. Both the
parabolic solution presented here and the Gaussian variational
ansatz indicate that above
$\varepsilon_{\mathrm{dd}}^{\mathrm{crit}}$ the system is liable
to collapse towards an infinitely thin and long prolate `pencil'
oriented along the field polarization direction i.e.\ $\kappa
\rightarrow 0$, since the system lowers its energy by arranging
the dipoles end-to-end. However, in reality a transition to
another (more structured?) state (see
\cite{goral00,giovanazzi2002a,goral2002,odell03,santos03})
presumably occurs in preference to the system becoming truly
singular. Bearing this in mind we shall consider below three
nominally different types of instability: local density
perturbations, `scaling' deformations, and the `Saturn-ring'
instability. The latter occurs due to a peculiarity in the
potential that would be seen by an atom located \emph{outside} the
boundary of the BEC, and may result in a previously unforeseen
type of instability. All of them predict the onset of instability
when $\varepsilon_{\mathrm{dd}} \ge 1$. Nevertheless, we have
included in the figures values of $\varepsilon_{\mathrm{dd}}$
exceeding unity, our justification being partly mathematical
curiosity, and partly the fact that the inclusion of kinetic
energy would extend the stability of a dipolar BEC beyond that of
the strict Thomas-Fermi limit considered here.

\subsection{Local density perturbations}
Phonons have already been predicted \cite{goral00,giovanazzi2002b}
to cause instabilities in a \emph{homogeneous} dipolar BEC when
$\varepsilon_{\mathrm{dd}}>1$. This can be seen directly from the
Bogoliubov dispersion relation between the energy $E_{\mathrm{B}}$
and momentum $p$ for phonons in the gas
\begin{equation}
E_{\mathrm{B}}=\sqrt{\left(\frac{p^{2}}{2m}\right)^{2}
+2gn\left\{1+\varepsilon_{\mathrm{dd}}
\left(3 \cos^{2} \theta -1 \right) \right\}\frac{p^{2}}{2m}}
\label{eq:bogdispersion}
\end{equation}
which can become imaginary when $\varepsilon_{\mathrm{dd}}>1$,
indicating an instability. This dispersion relation has an angular
dependence ($\theta$ is the angle between the momentum of the
phonon and the external polarizing field) which further
illustrates the richness of dipolar systems in comparison to the
usual non-dipolar case. Equation (\ref{eq:bogdispersion}) is
derived by adding to the Fourier transform, $g$, of the contact
interaction $g\delta(\br)$ that appears in the usual Bogoliubov
dispersion relation, the Fourier transform, $C_{\mathrm{dd}}
(\hat{k}_{i}\hat{k}_{j}-\delta_{ij}/3 )$, of the dipole-dipole
interaction (\ref{eq:staticdipdip}). This is an approximation that
assumes, as we do throughout this paper, that there is no
screening of two dipoles by the other dipoles lying between them
and also that the scattering of these two particles by the
dipole-dipole interaction takes place within the Born
approximation, as mentioned above in Sec. \ref{sec:TFeqn}.

In a trapped BEC with negligible kinetic energy, as considered here, we should
expect an analogous instability due to local density perturbations
having a wavelength much smaller than the dimensions of the
condensate. For example, Santos \emph{et al} \cite{santos03}
recently showed that an infinite-pancake dipolar BEC, homogenous
in two directions and parabolic in the third, is unstable when
$\varepsilon_{\mathrm{dd}}>1$ for a density exceeding a critical
value.

\subsection{Scaling deformations}
We use the term `scaling' deformation to describe perturbations
that merely re-scale the parabolic solution
(\ref{eq:densityprofile}), i.e.\ that change $R_{x}=R_{y}$ and
$R_{z}$ from their equilibrium values (\ref{eq:Rxsol}), but leave
the basic form of the parabolic solution the same. Since the
equilibrium values of the radii are determined by the
transcendental equation (\ref{eq:transcendental}), which also
occurs in the context of a Gaussian variational solution, much of
what we shall say below has already been described by other
authors, including Santos \emph{et al} \cite{santos00}, and Yi and
You \cite{yi01}.

\begin{figure}[t]
\vspace*{-4mm}
\begin{center}
\centerline{\epsfig{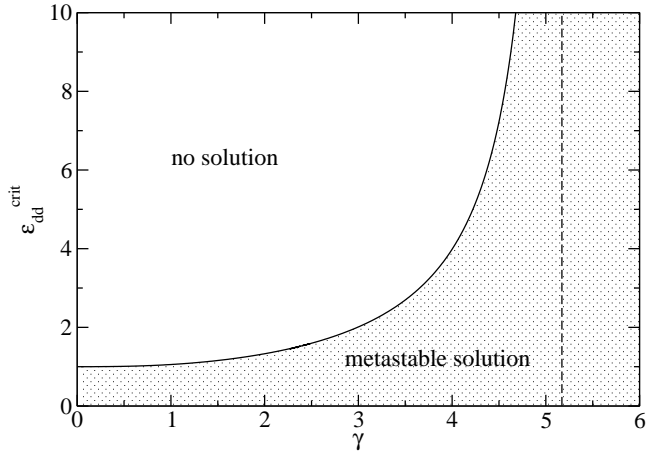}}
\end{center}
\vspace*{-5mm} \caption{The critical value of the dipolar
coupling, $\varepsilon_{\mathrm{dd}}^{\mathrm{crit}}$, above which
the condensate becomes strictly unstable---even the  metastable
state (local minimum in the energy landscape) no longer exists.
However, above $\gamma_{\mathrm{crit}}=5.1701$, there is always a
solution metastable to scaling deformations even for arbitrarily
large $\varepsilon_{\mathrm{dd}}$ (although not necessarily stable
to local density perturbations---see text).} \label{fig:ecrit}
\end{figure}

Information on the stability of the parabolic solution can be
gained from analyzing the behavior of the energy functional
$E_{\mathrm{tot}}=E_{\mathrm{trap}}+E_{s\mathrm{-wave}}+E_{\mathrm{dd}}$
evaluated over a general parabolic density profile
(\ref{eq:densityprofile})
\begin{equation}
E_{\mathrm{tot}}= \frac{N}{14} m \omega_{x}^{2} R_{x}^{2}
\left(2+\frac{\gamma^{2}}{\kappa^{2}} \right) + \frac{15}{28 \pi}
\frac{N^{2}g}{R_{x}^{2}R_{z}}\left[1-
\varepsilon_{\mathrm{dd}}f(\kappa) \right]
\label{eq:energyfunctional}
\end{equation}
in the vicinity of the solution (\ref{eq:Rxsol}) and
(\ref{eq:transcendental}) and across the whole parameter space
($\kappa,\varepsilon_{\mathrm{dd}},\gamma$). One finds that for $0
\le \varepsilon_{\mathrm{dd}} <1$ the solution given by the
transcendental equation (\ref{eq:transcendental}) is always
stable, in the sense that it corresponds to a global minimum of
the energy functional. As $\varepsilon_{\mathrm{dd}}$ is increased
and passes through unity, the solution matches smoothly onto one
that is only metastable, i.e.\ it is only a local minimum in the
energy landscape, and the global minimum is then a (prolate)
collapsed state with $\kappa \rightarrow 0$. Simultaneously with
the turning of the stable into a metastable solution, a second
branch of solutions appear at a smaller value of $\kappa$, which
correspond to unstable saddle points that separate the local
minimum of the metastable solution from the global minimum of the
collapsed state at $\kappa=0$ in the energy landscape. If one
continues to increase $\varepsilon_{\mathrm{dd}}$ then one of two
things happens, depending on the value of $\gamma$: if $\gamma$ is
less than a critical value, $\gamma < \gamma_{\mathrm{crit}} =
5.1701$, then, as $\varepsilon_{\mathrm{dd}}$ increases,
eventually the metastable and unstable solutions coalesce at
$\varepsilon_{\mathrm{dd}}=\varepsilon_{\mathrm{dd}}^{\mathrm{crit}}$,
above which there are no solutions, not even metastable ones, and
the energy landscape is just a continuous slope down towards a
collapsed state with $\kappa=0$. This critical value
$\varepsilon_{\mathrm{dd}}^{\mathrm{crit}}$ as a function of
$\gamma$ is plotted in Figure \ref{fig:ecrit}. If, however,
$\gamma>\gamma_{\mathrm{crit}}$ then something rather surprising
happens. As first remarked by Santos \emph{et al} \cite{santos00},
when $\gamma>\gamma_{\mathrm{crit}}$ there exists a solution
metastable to scaling perturbations at a finite value of $\kappa$
for all values of $\varepsilon_{\mathrm{dd}}$, strictly speaking
even for $\varepsilon_{\mathrm{dd}}= \infty$. (Note, however, that
our value for $\gamma_{\mathrm{crit}}$ disagrees with that of
Ref.~\cite{santos00} but is that same as the one given in
Ref.~\cite{yi01}).

For completeness we would like to mention that prima facie the
transcendental equation (\ref{eq:transcendental}) for
$\varepsilon_{\mathrm{dd}}>1$ has solutions also for
$\kappa>\gamma$. These come in pairs, one corresponding to a
maximum and the other to a saddle point in the energy landscape.
However, inspection reveals that these solutions have no physical
relevance as for them the radius $R_{x}$ of the condensate,
Eq.~(\ref{eq:Rxsol}), comes out imaginary.

\subsection{Saturn-ring instability} \label{subsec:saturn}
An examination of the dipolar potential outside the condensate,
that is seen, for example by a single test atom placed beyond the
boundary $\rho^{2}/R_{x}^{2}+z^{2}/R_{z}^{2}=1$, reveals a new
type of possible instability. Like the local density
perturbations, it also does not preserve the parabolic form of the
density profile.  It turns out that for
$\varepsilon_{\mathrm{dd}}>1$ the potential seen by atoms just
outside the condensate exhibits a local minimum, i.e. the sum of
trap and dipole-dipole potentials is locally lower than the
chemical potential, which causes atoms to spill out from the
condensate and fill this dip in the potential. Such an effect is
peculiar to condensates with induced dipole-dipole interactions
because these are long-range and thus give rise to a potential
even outside the condensate, whereas the potential due to $s$-wave
scattering is short-range and thus zero outside the condensate. To
investigate the dip in the outside potential we need to use the
dipolar potential (\ref{eq:outprolate},\ref{eq:outoblate}) that we
calculated in Section \ref{sec:outside}. As the expressions are
rather awkward in Cartesian coordinates, we shall analyze them in
spheroidal coordinates.

The outside potential ${\cal V}$ is the sum of the trap potential
and the dipole-dipole interaction potential
$\Phi_{\mathrm{dd}}^{\rm (outside)}({\bf r})$. The trap potential
is positive and monotonically rising from the center. The
dipole-dipole interaction potential outside the condensate is a
solution of the (homogeneous) Laplace equation, i.e.\
$\Phi_{\mathrm{dd}}^{\rm (outside)}({\bf r})$ is a harmonic
function. Thus the Maximum Principle applies, and
$\Phi_{\mathrm{dd}}^{\rm (outside)}({\bf r})$ must assume its
maximum and minimum on the boundaries of the domain, either at
infinity or on the surface of the condensate. At infinity the
dipole-dipole potential vanishes, which means that at large
distances the total outside potential is positive and dominated by
the trap potential. To ascertain whether the outer potential
${\cal V}$ has a local minimum one only needs to check whether the
sum of trap and dipole-dipole potentials has a negative first
derivative at the surface of the condensate in some outward
direction. It is easy to see that local minima of ${\cal V}$ can
occur only for $\eta=0$: at a local minimum of ${\cal V}$ its
first derivative with respect to $\xi$ and $\eta$ must vanish, but
both the trap potential and the dipole-dipole potential are
quadratic in $\eta$, so that the first derivative $\partial{\cal
V}/\partial \eta$ is proportional to $\eta$ and thus vanishes only
for $\eta=0$ unless it vanishes for all $\eta$.
\begin{figure}[t]
\vspace*{-12pt}
\begin{center}
\centerline{\epsfig{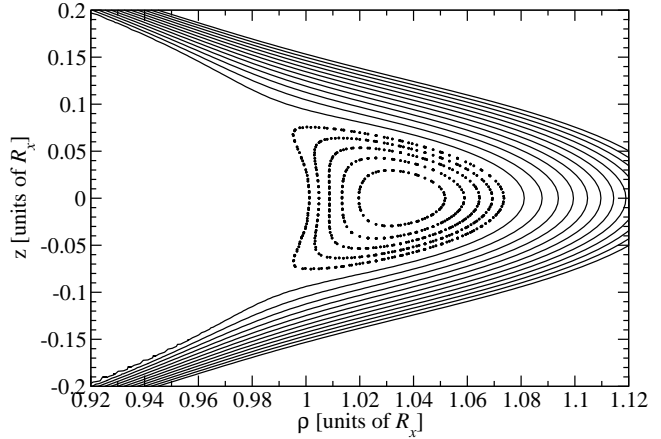}}
\end{center}
\vspace*{-5mm} \caption{Contour plot of the potential ${\cal V}$
(in units of the chemical potential $\mu$) outside a condensate
with $\kappa=2$ at $\varepsilon_{\mathrm{dd}}=1.5$. The closed
contours drawn in dotted lines have ${\cal V}/\mu<1$; the lowest
in the middle is at ${\cal V}/\mu=0.992$ and the difference to the
next higher is roughly 0.0017. The cut off contours outside the
condensate go from ${\cal V}/\mu=1.002$ to 1.04 in steps of
approximately 0.0034.} \label{fig:dip}
\end{figure}
Therefore we only need to examine the derivative
$\partial{\cal V}/\partial\xi$ at $\eta=0$; we find
\begin{equation}
\frac{\partial}{\partial\xi} \left.\left( \frac{\cal V}{\mu}
\right)\right|_{\xi=\xi_B, \eta=0} =
\frac{2(1-\varepsilon_{\mathrm{dd}})}{\xi_B\kappa^2
[1-\varepsilon_{\mathrm{dd}}f(\kappa)]}
\label{eq:firstderiv}
\end{equation}
where $\xi_B$ is the value of the spheroidal variable $\xi$ on the
surface of the condensate, i.e.\ $\xi_B=1/\sqrt{1-\kappa^2}$ for a
prolate BEC and $\xi_B=1/\sqrt{\kappa^2-1}$ for an oblate BEC. For
oblate BECs the denominator in Eq.\ (\ref{eq:firstderiv}) is
always positive, and thus $\partial{\cal V}/\partial\xi$ is
negative if and only if $\varepsilon_{\rm dd}>1$. For prolate BECs
the denominator is always positive in the stable and metastable
regions of Fig \ref{fig:aspect}, and thus in these regions
$\partial{\cal V}/\partial\xi$ is negative if and only if
$\varepsilon_{\rm dd}>1$. Whenever $\partial{\cal V}/\partial\xi$
is negative on the condensate surface, a local minimum of the
potential lies somewhere outside the surface. The fact that this
happens at $\eta=0$ means that the local dip in the potential
occurs along a ring at $z=0$ around the condensate, and the
flowing out of atoms from the main condensate into the dip causes
the condensate to take on a Saturn-like appearance, which then
leads to further instability. Figure \ref{fig:dip} gives an
illustration of a typical potential, plotted as contours in the
$\rho-z$ plane. The flat part to the left of the center is the
constant chemical potential inside the condensate.

\section{Electrostriction}
\begin{figure}[tp]
\vspace*{-4mm}
\begin{center}
\centerline{\epsfig{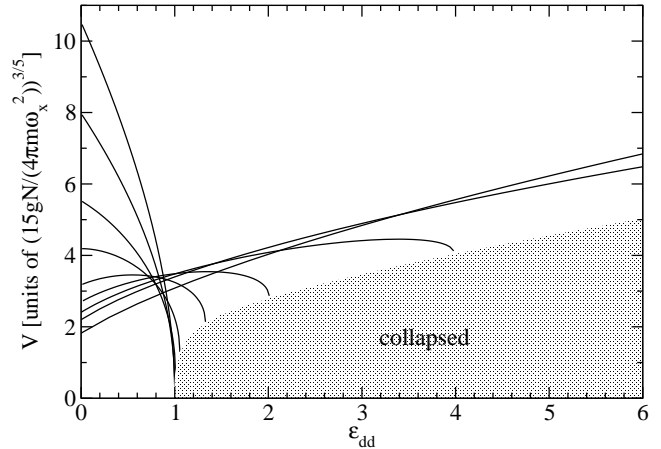}}
\end{center}
\vspace*{-5mm} \caption{The volume V of the condensate for traps
with aspect ratios $\gamma$=\{0.1,0.2,0.5,1,2,3,4,5,8\} as a
function of the dipole-dipole coupling strength
$\varepsilon_{\mathrm{dd}}$. At $\varepsilon_{\mathrm{dd}}=0$,
i.e.~without dipole-dipole interactions, the volume is
proportional to $\gamma^{-2/5}$, so that on the left edge of the
graph higher volume corresponds to lower $\gamma$. That reverses
with increasing $\varepsilon_{\mathrm{dd}}$. } \label{fig:volume}
\end{figure}

The volume $V$ of the spheroidal BEC can be expressed as
\begin{equation}
V=\frac{4 \pi}{3} R_{x}^{2} R_{z} = \frac{4 \pi}{3}
\frac{R_{x}^{3}}{\kappa} \;.
\end{equation}
Substituting $R_{x}$ from Eq.~(\ref{eq:Rxsol}) we can write it as
\begin{eqnarray}
V = \frac{4 \pi}{3 \kappa^{2/5}}  \left[\frac{15 g N}{4 \pi m
\omega_{x}^{2}} \right]^{3/5} \left[1+ \varepsilon_{\mathrm{dd}}
\left( \frac{3}{2} \frac{\kappa^{2} f(\kappa)}{1-\kappa^{2}}-1
\right)
  \right]^{3/5}
\end{eqnarray}
and then we can use the transcendental equation
(\ref{eq:transcendental}) to eliminate $\kappa$ in favor of
$\gamma$ and $\varepsilon_{\mathrm{dd}}$. In Fig.~\ref{fig:volume}
we plot $V$ in units of $[15gN/(4 \pi m\omega_{x}^2)]^{3/5}$ as a
function of $\varepsilon_{\mathrm{dd}}$ for various trap aspect
ratios $\gamma$. The figure shows that condensates in prolate
traps and slightly oblate traps (with $\gamma<1.6630$) get
compressed by increasing dipolar interactions, while condensates
in traps with aspect ratios above $\gamma_{\mathrm{crit}}=5.1701$
are being pulled apart as $\varepsilon_{\mathrm{dd}}$ rises.
Between $\gamma=1.6630$ and $\gamma_{\mathrm{crit}}$ is a range of
trap aspect ratios for which the condensate is pulled apart at
first but eventually compressed into collapse by higher values of
the dipolar interaction strength. If, during an experiment, the
condensate was imaged in the trap then the volume could be
measured either directly from the radii, or by the central density
which is inversely proportional to the volume, $V=(5/2)N/n_{0}$.

\section{Release Energy}
If the trap is turned off the condensate expands ballistically and
the $s$-wave and dipole-dipole interaction energies are converted
into kinetic energy, the so-called release energy, which can be
measured in an experiment. For the exact parabolic solution the
release energy is given by
\begin{equation}
E_{\mathrm{rel}}=15 g
N^{2}[1-\varepsilon_{\mathrm{dd}}f(\kappa)]/(28 \pi R_{x}^{2}
R_{z}).
\end{equation}
The release energy can also be expressed in terms of the chemical
potential (\ref{eq:chemical_potential}) as $E_{\mathrm{rel}}=(2/7)
N \mu$, which is the standard expression for the Thomas-Fermi
limit \cite{dalfovo99}.

\section{Conclusions}
The long-range anisotropic nature of dipole-dipole interactions
gives condensates composed of dipolar atoms or molecules novel
properties compared to those with only short-range interactions.
Furthermore, the dipole-dipole interactions can be controlled in
sign, magnitude and direction. We have shown that a simple
inverted parabola remains an exact solution for the
density-profile of a harmonically trapped dipolar BEC in the
Thomas-Fermi limit by calculating the dipolar potential both
inside and outside the condensate region. The parabolic solution
is stable, in the strict Thomas-Fermi limit, only for
$\varepsilon_{\rm dd}<1$. For $\varepsilon_{\rm dd}>1$ it is
unstable against scaling perturbations as seen in the energy
functional, against perturbations of different symmetry, such as
phonons, and also due to a local minimum appearing in the
potential outside the condensate. The effect of the dipole-dipole
interactions upon the BEC is to change both its aspect ratio and
its volume. The manner of this electrostriction depends on the
aspect ratio of the external trap; as a function of the strength
of the dipole-dipole interactions (relative to the $s$-wave
scattering) the volume of the BEC can either increase or decrease,
or even initially increase and then decrease at higher values of
the coupling.

\begin{acknowledgments}
It is a pleasure to thank Gabriel Barton and Tilman Pfau for
discussions. We would like to acknowledge financial support from
the UK Engineering and Physical Sciences Research Council (D.O'D.)
and from the European Union (S.G.), and we are indebted to the
Royal Society for a University Research Fellowship (C.E.).
\end{acknowledgments}

\appendix
\section{Integrating over homoeoid shells}
\label{appendixA} Here we demonstrate how the potential
(\ref{eq:resultphi}) can be obtained by integrating over
successive thin shells, a method that works even in the general
case of an ellipsoid. The problem of the
electrostatic/gravitational field within and outside a
charged/massive ellipsoid is associated with some of the great
names of nineteenth century mathematical physics. In his 1892
treatise on statics, Routh \cite{routh} notes that Chasles,
Dirichlet, Jacobi, and Poisson all made contributions. Rodrigues
is credited with being the first to evaluate the potential of a
solid ellipsoid of constant density, seemingly in 1815, and in
1833 Green \cite{green} ``treated the subject in a very general
manner'' giving solutions for various cases of inhomogeneous
density, as in the current situation. Our derivation is adapted
from Routh. Although we take a cylindrically symmetric density as
input, the derivation holds for the general ellipsoidal case.

Consider an ellipsoidal surface
$(x/R_{x})^{2}+(y/R_{y})^2+(z/R_{z})^{2}=1$, having semi-axes
($R_{x},R_{y},R_{z}$). A continuous family of concentric
\textit{similar} ellipsoids lying inside this outer surface can
then be defined via the semi-axes ($s R_{x}, s  R_{y}, s  R_{z}$),
where $0\le s \le 1$. Note that by similar we mean that this
family all share the same aspect ratios among their semi-axes, but
are consequently \emph{not} confocal. A surface which is
\emph{confocal} to the original surface obeys
$x^{2}/(R_{x}^{2}+\lambda)+y^{2}/(R_{y}^{2}+\lambda)
+z^{2}/(R_{z}^{2}+\lambda)=1$,
so that $\lambda=0$ is obviously the original surface, and
surfaces with $\lambda>0$ lie outside the original. Considered in
terms of the parameterization specified by $s$, the parabolic
density profile can be written as $n=n_{0}(1-s^{2})$. Since the
equi-density surfaces within the density profile are similar, when
computing the integral (\ref{eq:electrostaticgreen2}) it makes
sense to take as our basic volume element a thin
``homoeoid''\cite{routh}, defined as the shell bounded by two
similar ellipsoidal surfaces, parameterized by $s$ and $s+\romand
s$, respectively. The volume of the thin homoeoid is $\romand V=4
\pi R_{x}R_{y}R_{z} s^{2}\romand s $. When computing the potential
at a point $\mathcal{P}(x,y,z)$ within a charged ellipsoid the
contribution from the homoeoids interior to that point is of a
different nature to that from those exterior to it, so we consider
the two parts separately.

\subsection{ Potential $\phi^{\mathrm{in}}$ due to `charge'
interior to $\mathcal{P}$}
Let $\romand\phi$ denote the
contribution to the total potential $\phi$ from a thin homoeoid
shell. In this section we require the potential $\romand
\phi^{\mathrm{in}}$ \emph{outside} a thin homoeoid, labelled by
its inner surface $s$, whose `charge' density is
$n=n_{0}(1-s^{2})$. The equi-potential surfaces outside a charged
thin homoeoid $s$ are confocal ellipsoids \cite{routh}
\begin{equation}
\frac{x^{2}}{s^{2}R_{x}^{2}+\lambda}+\frac{y^{2}}{s^{2}R_{y}^{2}+\lambda}
+\frac{z^{2}}{s^{2}R_{z}^{2}+\lambda}=1 \label{eq:confocalsurface}
\end{equation}
In our analogy (in which the dielectric constant $\epsilon_{0}=1$)
the `potential' on a confocal surface $\lambda$ outside a thin
homoeoid labelled by $s$, of density $n$ and volume $\romand V$ is
\begin{equation}
\romand\phi^{\mathrm{in}}_{\lambda}(s) =  \frac{n \; \romand V}{8
\pi } \int_{\lambda}^{\infty}\!\! \frac{\romand
u}{\sqrt{(s^{2}R_{x}^{2}+u)(s^{2}R_{y}^{2}+u)(s^{2}R_{z}^{2}+u)}}
 \label{eq:homoeoidpotential}.
\end{equation}
This potential is most easily obtained by noticing that the
`charge' distribution on the homoeoid is identical to that of a
solid ellipsoidal \emph{conductor} with the same external
dimensions and with the same total `charge' ($n\romand V$)
distributed over its surface. A homoeoid shell does not have
uniform thickness, $\romand h$, being slightly thicker at the
points furthest from the center
\begin{equation}
\romand h=\frac{s \;\romand
s}{\displaystyle\sqrt{\frac{x^{2}}{R_{x}^{4}}
+\frac{y^{2}}{R_{y}^{4}}+\frac{z^{2}}{R_{z}^{4}}}}.
\end{equation}
If, rather than a shell of thickness $\romand h$ we consider the
`charge' it contains to in fact be surface charge, of variable
surface density $\omega(x,y,z)$, i.e.\ so that $\omega=n \romand
h$, we obtain exactly the same $\omega(x,y,z)$ as in the
well-known problem of the charged ellipsoid conductor (see, e.g.\
\cite{stratton}), and thus the resulting potentials are also the
same.

Having established the potential due to a homoeoid shell we must
integrate over all the shells $s$ lying inside $\mathcal{P}$,
which is located on the confocal surface $\lambda$. While
$\mathcal{P}$ is of course fixed in space, the surface $\lambda$
passing through it is a different surface (i.e.\ different aspect
ratio) for each homoeoid, i.e. $\lambda=\lambda(s)$, becoming at
the last instant a similar surface to the final homoeoid (upon
which $\mathcal{P}$ itself sits). To simplify the algebra we put
$\lambda=s^{2} \sigma$ in the equation for the ellipsoidal surface
(\ref{eq:confocalsurface}). We see that if $s=0$ then
$\sigma=\infty$. We choose $s=\tilde{s}$ to describe the similar
ellipsoidal surface upon which $\mathcal{P}$ lies, and thus we
have $\sigma=0$ for $s=\tilde{s}$. To obtain the potential due to
the charge interior to $\mathcal{P}$ we must evaluate
$\phi^{\mathrm{in}}=\int_{s=0}^{\tilde{s}} \romand
\phi^{\mathrm{in}}_{\lambda}(s)$. Setting $u=s^{2}v$ in the
integral (\ref{eq:homoeoidpotential}), we find
\begin{eqnarray}
\displaystyle \phi^{\mathrm{in}} & = & \frac{n_{0}
R_{x}R_{y}R_{z}}{2}\! \times  \\ & & \int_{0}^{\tilde{s}}\!\!
\romand s \; (1-s^{2}) \: s \int_{\sigma(s)}^{\infty}\!
\frac{\romand v}{\sqrt{(R_{x}^{2}+v)(R_{y}^{2}+v)(R_{z}^{2}+v)}}
\nonumber
\end{eqnarray}
which can be integrated by parts to give
\begin{eqnarray}\displaystyle
\phi^{\mathrm{in}} & = & \frac{n_{0} R_{x}R_{y}R_{z}}{2} \bigg\{
\nonumber \\ & &
\left(\frac{\tilde{s}^{2}}{2}-\frac{\tilde{s}^{4}}{4} \right)
\int_{0}^{\infty}\!\!\frac{\romand
v}{\sqrt{(R_{x}^{2}+v)(R_{y}^{2}+v) (R_{z}^{2}+v)}}
\nonumber \\
&&  - \int_{0}^{\infty} \left[\frac{1}{2}
\left(\frac{x^{2}}{R_{x}^{2}+\sigma}+\frac{y^{2}}{R_{y}^{2}+\sigma}
+\frac{z^{2}}{R_{z}^{2}+\sigma}\right)\right.\nonumber \\
&& \left.- \frac{1}{4}
\left(\frac{x^{2}}{R_{x}^{2}+\sigma}+\frac{y^{2}}{R_{y}^{2}+\sigma}
+\frac{z^{2}}{R_{z}^{2}+\sigma}\right)^{2} \right]\nonumber\\
&&\left.\times\frac{\romand \sigma}{\sqrt{(R_{x}^{2}+\sigma)
(R_{y}^{2}+\sigma)(R_{z}^{2}+\sigma)}}\right\}\;,
\label{eq:homoeoidpotentialinterior}
\end{eqnarray}
where in the second term we have used Leibniz's formula for
differentiating an integral and used Eq.\
(\ref{eq:confocalsurface}) to replace $s^{2}$ and $s^{4}$.

\subsection{Potential $\phi^{\mathrm{ex}}$ due to `charge' exterior to
$\mathcal{P}$}
The potential inside a homoeoid is much simpler to
calculate since, just as in the case of the spherical shell, it is
a constant, independent of position \cite{portis}. Returning to
the solid ellipsoidal conductor model, the potential throughout a
conductor is the same as on the surface defined by $\lambda=0$.
Thus
\begin{displaymath}
\romand\phi^{\mathrm{ex}}(s) =  \frac{n \; \romand V}{8 \pi }
\int_{0}^{\infty} \frac{\romand
u}{\sqrt{(s^{2}R_{x}^{2}+u)(s^{2}R_{y}^{2}+u)(s^{2}R_{z}^{2}+u)}}.
\end{displaymath}
Note that because the lower limit of the integral is this time
independent of $s$ the ensuing treatment is simple. Integrating
from the surface $\tilde{s}$, which includes $\mathcal{P}$, out
the boundary $s=1$, we require
$\phi^{\mathrm{ex}}=\int_{s=\tilde{s}}^{1} \romand
\phi^{\mathrm{ex}}(s)$. Making once again the substitution
$u=s^{2}v$ one immediately finds
\begin{eqnarray}
\displaystyle \phi^{\mathrm{ex}} &&=
\frac{n_{0}R_{x}R_{y}R_{z}}{2}\left[ \frac{1}{2}(1-\tilde{s}^{2})
-\frac{1}{4}(1-\tilde{s}^{4}) \right]
\nonumber \\
&& \times \int_{0}^{\infty}\frac{\romand
v}{\sqrt{(R_{x}^{2}+v)(R_{y}^{2}+v)(R_{z}^{2}+v)}}\;.
\label{eq:homoeoidpotentialexterior}
\end{eqnarray}

Combining Eqs.~\ref{eq:homoeoidpotentialinterior} and
\ref{eq:homoeoidpotentialexterior} we see that the parameter
$\tilde{s}$ drops from the sum
$\phi^{\mathrm{tot}}=\phi^{\mathrm{in}} + \phi^{\mathrm{ex}}$.  In
the general ellipsoidal case the remaining integrals in
$\phi^{\mathrm{tot}}$ can be expressed in terms of elliptic
integrals. In the spheroidal case ($R_{x}=R_{y}$) they can be
written in terms of more elementary functions. There are two cases
to distinguish, depending upon whether the density profile is
prolate (a cigar), $R_{z}>R_{x}$, or oblate (a pancake),
$R_{x}>R_{z}$. If we define
\begin{eqnarray}
 \mathcal{J}(a,c) & \equiv & \int_{0}^{\infty}
\frac{\romand v}{(R_{x}^{2}+v)\sqrt{(R_{z}^{2}+v)}}
\nonumber \\
 & = & \left\{ \begin{array}{ll}\displaystyle
\frac{1}{\sqrt{R_{z}^{2}-R_{x}^{2}}}
\ln\left(\frac{1+\sqrt{1-R_{x}^{2}/R_{z}^{2}}}{1-\sqrt{1-R_{x}^{2}
/R_{z}^{2}}}\right)
& \mathrm{prolate} \\\displaystyle \frac{2}{
\sqrt{R_{x}^{2}-R_{z}^{2}}} \arctan
\left(\sqrt{R_{x}^{2}/R_{z}^{2}-1} \right) & \mathrm{oblate}
\end{array} \right.\nonumber
\end{eqnarray}
then the total potential at $\mathcal{P}$ is most compactly
expressed as
\begin{eqnarray}\displaystyle
\phi^{\mathrm{tot}} && =  \frac{n_{0} R_{x}^{2}R_{z}}{2} \left(
\frac{\mathcal{J}}{4}+\frac{\rho^{2}}{2} \frac{\partial {\cal
J}}{\partial (R_{x}^{2})}+z^{2}\frac{\partial
\mathcal{J}}{\partial (R_{z}^{2})} \right.
\label{eq:totalpotentialhomoeoid} \\
&& \left.+\frac{\rho^{4}}{8}\frac{\partial^{2}
\mathcal{J}}{\partial (R_{x}^{2})^{2}}
+\frac{z^{4}}{3}\frac{\partial^{2} \mathcal{J}}{\partial
(R_{z}^{2})^{2}}+ \rho^{2}z^{2}\frac{\partial^{2} \mathcal{
J}}{\partial (R_{x}^{2})\partial (R_{z}^{2})} \right)\; \nonumber
\end{eqnarray}
and is identical to the Green's function result
(\ref{eq:resultphi}). The relationship between the function $\Xi$
defined in the Green's function approach and $\mathcal{J}$ defined
here is simply $\mathcal{J}= \Xi/R_{z}$.

\section{The potential outside the condensate}
\label{appendixB} The calculation of the potential at a point
$\mathcal{P}$ outside the boundary of the condensate follows very
similar lines to that presented above for a point inside. The main
difference is that one no longer has to consider `charge' located
exterior to $\mathcal{P}$. The result is identical to Eq.\
(\ref{eq:totalpotentialhomoeoid}) except that the integral
$\mathcal{J}$ is now given by
\begin{widetext}
\begin{eqnarray}
 {\cal J}(a,c,\lambda) &\equiv & \int_{\lambda}^{\infty}
\frac{\romand v}{(R_{x}^{2}+v)\sqrt{(R_{z}^{2}+v)}}
= \left\{ \begin{array}{ll}\displaystyle
\frac{1}{\sqrt{R_{z}^{2}-R_{a}^{2}}}
\ln\left(\frac{\sqrt{\lambda+R_{z}^{2}}+
\sqrt{R_{z}^{2}-R_{x}^{2}}}{\sqrt{\lambda+R_{z}^{2}}
-\sqrt{R_{z}^{2}-R_{x}^{2}}}\right)
& \mathrm{prolate} \\\displaystyle \frac{2}{
\sqrt{R_{x}^{2}-R_{z}^{2}}} \arctan
\left(\sqrt{\frac{R_{x}^{2}-R_{z}^{2}}{\lambda+R_{z}^{2}}} \right)
& \mbox{oblate}
\end{array} \right.\nonumber
\end{eqnarray}
where $\lambda$ parameterizes an ellipse, confocal to the
outer-boundary of the condensate, upon which the point
$\mathcal{P}$ sits, and so obeys the quadratic equation
$\rho^{2}/(R_{x}^{2}+\lambda)+z^{2}/(R_{z}^{2}+\lambda)=1$. When
solving for $\lambda$ one should take the positive solution
\begin{displaymath}
\lambda = \frac{\rho^{2}-R_{x}^{2}+z^{2}-R_{z}^{2}}{2}+
\frac{\sqrt{\left(\rho^{2}-R_{x}^{2}+z^{2}-R_{z}^{2}\right)^{2}+4
\left(\rho^{2}R_{z}^{2}+z^{2}R_{x}^{2}-R_{x}^{2}R_{z}^{2}\right)}}{2}.
\end{displaymath}
\end{widetext}
Note that expressing the solution for the potential $\phi$ in the
form (\ref{eq:totalpotentialhomoeoid}) requires that when taking
the derivatives with respect to $R_{x}$ and $R_{z}$ then $\lambda$
is to be treated as a constant and is \emph{not} to be
differentiated. On the other hand, when going on to calculate the
dipolar potential outside the BEC, $\Phi_{\mathrm{dd}}=
-C_{\mathrm{dd}}\: \hat{\mathrm{e}}_{i} \hat{\mathrm{e}}_{j}
\nabla_{i} \nabla_{j} \phi(\br)$, then $\lambda$ \emph{is} to be
treated as a function of $(\rho,z)$, and should be differentiated.
This makes the exact expression for $\Phi_{\mathrm{dd}}$ outside
the condensate complicated. However, in the limit $\rho \gg
R_{x}$, $z \gg R_{z}$, one may use the asymptotic result
\begin{equation}
\phi^{\mathrm{outside}} \sim \frac{N}{8 \pi}
\frac{(R_{x}^{2}-R_{z}^{2})(\rho^{2}-2z^{2})+14(\rho^{2}+z^{2})^{2}}{7(
\rho^{2}+z^{2})^{5/2}}
\end{equation}
which turns out in practice to be remarkably accurate, even right
up to the condensate surface providing the condensate is not too
aspherical (in which case the next terms in the multipole
expansion should included.)


\begin{thebibliography}{99}
\bibitem{mistake} Note that eq. (10.3.63) in \cite{morse} is misprinted and
should be divided by 2.
\end{thebibliography}
\end{document}